# Toward III-V/Si co-integration by controlling biatomic steps on hydrogenated Si(001)

M. Martin[1], D. Caliste[2], R. Cipro[1,3], R. Alcotte[1,3], J. Moeyaert[1], S. David[1], F. Bassani[1], T. Cerba[1,3], Y. Bogumilowicz[3], E. Sanchez[4], Z. Ye[4], X.Y. Bao[4], J.B. Pin[4], T. Baron[1] and P. Pochet[2,*]

[1]**Univ. Grenoble Alpes, CNRS-LTM, F-38054 Grenoble, France**

[2]**Univ. Grenoble Alpes, F-38000 Grenoble, France**

**CEA, INAC-MEM, F-38054, Grenoble, France**

[3]**Univ. Grenoble Alpes, F-38000 Grenoble, France**
**CEA, LETI, MINATEC Campus, F-38054 Grenoble, France**

[4]**Applied Materials, 3050 Bowers Avenue, Santa Clara, CA 95054, USA**

**Abstract**

The integration of III-V on silicon is still a hot topic as it will open up a way to co-integrate Si CMOS logic with photonic devices. To reach this aim, several hurdles should be solved, and more particularly the generation of antiphase boundaries (APBs) at the III-V/Si(001) interface. Density functional theory (DFT) has been used to demonstrate the existence of a double-layer steps on nominal Si(001) which is formed during annealing under proper hydrogen chemical potential. This phenomenon could be explained by the formation of dimer vacancy lines which could be responsible for the preferential and selective etching of one type of step leading to the double step surface creation. To check this hypothesis, different experiments have been carried in an industrial 300 mm MOCVD where the total pressure during the anneal step of Si(001) surface has been varied. Under optimized conditions, an APBs-free GaAs layer was grown on a nominal Si(001) surface paving the way for III-V integration on silicon industrial platform.

The introduction of new technological nodes in microelectronics devices is slowing down compared to the famous Moore's law. The introduction of three dimensional architectures, new materials and new functionalities in the vicinity of the logic core are key points to continue the improvement of integrated circuits performances. In the last decade, a regain of interest on the heterogeneous integration of III-V materials on silicon has been observed. Such combination allows to benefit from both the maturity of the silicon manufacturing technology and the III-V material electronic and optical properties (high electron mobility, direct band gap…). A wide variety of optoelectronic devices on a silicon platform has been demonstrated : high yield multijunction solar cells[1,2], high mobility n-Metal Oxide Semiconductor Field Effect Transistor(nMOSFET)[3-8] and also light sources and photodetectors[9,10]. The common way to integrate III-V on a Si platform is obtained via the transfer method applied to layers or dies[11,12]. Recently, thanks to hetero-epitaxy improvements, III-V monolithic integration on Si(001) becomes feasible. However, there are still challenges to be solved arising from the lattice mismatch and polarity differences especially. Indeed, the growth of zinc-blende structures, III-As and III-P polar films on a non-polar Si(001) substrate can lead to planar defects named antiphase boundary (APB). The APB planes are made of III-III or/and V-V bonds that can propagate in the layer through the {110}, {111} or higher index planes[13,14]. The elastic strain field due to APB changes atomic distances and hence electronic states, acting as a carrier diffusion and/or non-radiative recombination centers. APBs nucleate at the edges of the single-layer (SL) steps present at nominal (001) silicon surfaces. Until now, the APBs formation was mainly inhibited by using (i) off-axis Si(001) substrates tilted by 4-6° towards [110] direction[15,16] or (ii) Si(211) substrates[17] where Si double-layer (DL) steps could be formed easily. However these wafers are not compatible with industrial Si CMOS standards which uses nominal Si(001) wafers, i.e. with a miscut angle lower than 0.5°. The best option to prevent the APBs nucleation would be to be able to promote double layer (DL) step formation on nominal Si(001) substrate as it is the case for off-axis wafers. However, considering the thermodynamical models, the DL steps formation on nominal Si(001) is predicted as highly unfavorable in both ultra-high Vacuum (UHV) and $H_2$ atmosphere. In UHV the relaxation of strain induced by dimerization of the (2×1)-Si(001) reconstruction promotes single step (SL) formation until a miscut angle lying in a range between 1 and 3° [18-21]. Under hydrogen atmosphere the monohydride-terminated Si surface modifies the surface energy, leading to a phase transition between SL and DL step for miscut angle even higher than in UHV[22,23].

In this paper, a breakthrough is presented as we propose a reproducible and Si CMOS compatible solution to get APBs-free GaAs thin layer (<150 nm) epitaxially grown on standard Si(001) nominal 300 mm substrates. We will show that our computational modeling based on density functional theory (DFT), explains a possible mechanism to get double step formation on nominal Si(001). This theoretical results will be used to fix the experimental conditions to promote DL steps formation through Si(001) surface annealing under hydrogen atmosphere. In the following sections, the notation for steps and domains of Si(001) surface will be done according to Chadi's nomenclature[18]: a Si(001) surface with DL steps is made of A-type (or B-type) monodomains having Si-dimers perpendicular (or parallel) to the downward step edge. In that case, the DL steps are named $D_A$ (or $D_B$). A Si(001) surface with SL steps is made of alternating A-type and B-type domains. The corresponding SL steps are named $S_A$ and $S_B$.

First, we have theoretically studied the effect of hydrogenation on the thermodynamic stability of dimer-vacancy lines and rows on the Si(001) surface. The removal of two neighboring silicon atoms from this surface creates the so-called single-dimer vacancy (SDV)[24]. Line defects on the surface can appear by aligning SDV together, either in lines, creating dimer-vacancy lines (DVL) or in rows (DVR)[25] as shown respectively in green and pink areas in Fig. 1a. We have chosen to base our study on a 2×2 reconstructed surface because it reduces the asymmetry in the lateral elastic treatment of DVR and DVL as compared to a 2×1 surface. We have performed density functional theory (DFT) calculations using the Perdew-Burke-Ernzerhof exchange and correlation functional[26] with the code BigDFT[27]. The core electronic states are treated within the pseudo-potential formalism, using the Hartwigsen-Goedecker-Hutter formulation in the fitting proposed by Krach[28]. We have used a grid spacing of ~ 0.41 Bohr to project the wavefunctions, as this value has already proven to give accurate enough results for supercells with silicon[29]. The silicon surface has been represented by an elongated super-cell in surface boundary conditions. The super-cell is oriented along [110] and [-110] in its *x* and *y* directions respectively, while its *z* direction is collinear to [001]. It corresponds to a volume of 7.73 × 30.92 × 14.10 Å and features 9 layers of non-reconstructed silicon atoms along the *z* direction. To allow the relaxation of any elastic deformations induced by the defects on the surface and the surface reconstruction itself, we have added a silicon buffer of 18 layers along the *z* axis, treated by Stillinger-Weber force field as described in the work of Béland *et al.*[30]. A Monkhorst-Pack[31] *k*-point mesh of 4×1×1 has been used to properly describe the Brillouin zone in every direction. Every structure has been relaxed using a FIRE algorithm[32] to decrease forces on every atom below $5.10^{-4}$ Ht.Bohr$^{-1}$.

Then to take into account experimental conditions used in an industrial MOCVD reactor, hydrogen presence is added in our calculations. Formation energies $E_f$ are then computed from the comparison in total energy of the defected system (DVL or DVR, hydrogenated or not) with a 2×2 reconstructed silicon surface in the same super-cell[30]. The surface itself is considered either bare or mono-hydrogenated. For the bare surface we obtain quite similar formation energies for bare DVL and DVR, being $E_f$ =2.98 eV and $E_f$ =2.63 eV respectively. The energy difference between the two defects slightly raises to 0.51 eV for the mono-hydrogenated surface. In both situation, the two defects present different relaxed geometries. For the DVL case, the two silicon atoms that were previously bonded by the surface atom are shifted closer together to compensate for the missing bond, bringing long range deformations in the first bulk layer in a direction perpendicular to the DVL orientation. In the DVR case, where the defect width is larger than in the DVL case, the silicon atoms of the first bulk layer can rearrange themselves as tilted dimers in a similar way than for the 2×2 reconstructed surface, but with a 90° orientation compared to the surface reconstruction.

Next, the bare line defects are modified by placing a single hydrogen atom on each silicon of the first bulk layer with a dangling bond (Fig. 1a), changing their geometry and their formation energies. Indeed, our DFT calculations show that for both defects and whatever the surface state, the geometry distortions of the bare defects are considerably reduced when the defect is hydrogenated. In the DVR case, for instance, the dimers in the line adopt a flat position instead of a tilted one. This reduction in elastic stress is key in the formation energy lowering of the line defects. As shown on Fig. 1b, for hydrogen rich conditions (right hand-side of the graph Fig.1b), the formation energies of both hydrogenated DVR (H-DVR) and DVL (H-DVL) are lower than for the bare defects. Moreover, two regimes can be observed whatever the surface state. One range for super-high H chemical potentials where the H-DVL is favored, and a medium range of H chemical potentials where the H-DVR takes prominence. It is worth to note that the gain in energy per dimer can be quite important (several eVs) when comparing the different ranges. This leads to think that the selectivity with respect to H chemical potential is quite strong.

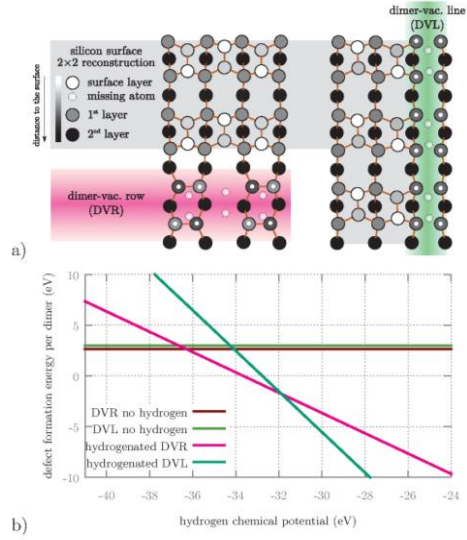

**Figure 1**: **(a)** schematic view of the DVR and DVL line defects on the 2×2 reconstructed silicon surface. Only two silicon bulk layers are represented (back and dark grey) in addition to the surface layer which is reconstructed. The distance from the surface is coded in gray-scale. Silicon atoms marked with a small white disk are hydrogenated in the case of hydrogenated DVR and DVL. **(b)** Represents the variation of the formation energy of both DVR and DVL defects, bare or hydrogenated with respect to the chemical potential of the hydrogen.

These results represent thermodynamic limits and can be used to discuss the stability of these various line defects on the Si(001) surface. For growth conditions without hydrogen, the DVR and DVL can co-exist (depending on the temperature) due to their close formation energies. Then, bringing hydrogen will have a strong selectivity impact in favor of H-DVR. Bringing still more hydrogen will reverse the selectivity for H-DVL. It is interesting to note that the geometry itself of the DVR can be seen as a precursor for steps on Si(001), see Fig. 1a. On the contrary, it is necessary to stack several DVL together to allow the silicon atoms of the first bulk layer to rearrange themselves into tilted dimers. As a consequence, we claim that it exists a range of hydrogen chemical potential where the creation of step on Si(001) is easier due to the H-DVR selectivity.

It is worth noticing that the reported hydrogen-induced bias between DVR and DVL is in the line of recent STM experiments[33]. Indeed, the observed selective etching of $S_B$ steps by hydrogen was identified to be driven by dimer-vacancies agglomeration into islands that progressively etch the steps. Interestingly the shape of the island is always elongated in the parallel direction for A-type domains and in the perpendicular direction for B-type domains, corresponding to DVR for both cases. The role of hydrogen is thus two fold. It first induces a large increase of dimer-vacancy concentration due to the lowering of their formation energy. While such increase could also be induced by Xe irradiation[34], the second effect of hydrogen is to select DVR with respect to DVL. This latter point is the key to explain the selective etching of $S_B$ steps. Indeed the steps themselves have an important factor ratio between the perpendicular and parallel directions which are respectively short and large. The correspondence of the former direction with DVR explains the progressive etching of $S_B$ steps as a consequence of the hydrogen-driven dimer-vacancy formation. This analysis helps to rationalize not only the double steps formation during the initial stage of hydrogen exposure but also the further layer by layer etching observed by Bruckner *et al.* (with the silicon surface oscillating between A-and B-type majority domains depending on the $H_2$ annealing time[33]). As a consequence, optimum DL steps should be only obtained in a restricted process window where dimer-vacancy concentration is large enough (driven by T and $P_{H2}$), with a significant bias between DVR wrt DVL (driven by $P_{H2}$) and for an optimized process time (to avoid the further etching of terraces with DL steps which would again lead to a surface with SL steps).

Next, we will use the above define surface engineering strategy. First we need to make a correspondence between the experimental $H_2$ pressure and the hydrogen chemical potential according to figure 1b. We use 300 mm diameter Czochralsky (CZ)-grown Silicon (001) wafers, sliced from the ingot with a slight 0.15° miscut angle towards either the [110] or [100] direction. Prior to the $H_2$ annealing, the native oxide is removed by SICONI[TM] process[35]. After transferring the wafers under vacuum, the $H_2$ annealing is performed directly in a 300 mm Applied Materials MOCVD reactor specifically designed for III-V materials growth and which uses purified $H_2$ as carrier gas. After the annealing process, the chamber is quickly cooled down to freeze the silicon surface step structure (30s to lower the temperature below 700°C). The morphology of the silicon surface was investigated by Atomic Force Microscope (AFM), in standard tapping mode in air. We have checked that the formation of native silicon dioxide doesn't prevent the observation of the surface structure.

The AFM image of figure 2a) shows the surface topography of a Si(001), with a 0.15° misorientation in the [110] direction, after a 10 min $H_2$ annealing at low pressure (in the range of 5-80Torr). The surface is made of terraces with SL steps, which is consistent with the result predicted by DFT at low pressure i.e. at low hydrogen chemical potential ($\mu_H$<36eV, left hand-side of the graph fig 1b). Indeed, the dimer vacancies generation is expected to be very low since the H-DVR/DVL formation energies are high, even higher than for bare defects. Such low vacancies density is not sufficient to remove the $S_B$ steps. Therefore the surface morphology is not kinetically driven by the vacancies generation and selective etching of $S_B$ steps but rather by the thermodynamic considerations for a nominal Si(001)-(2×1) surface, which means a surface with SL steps[22,23].

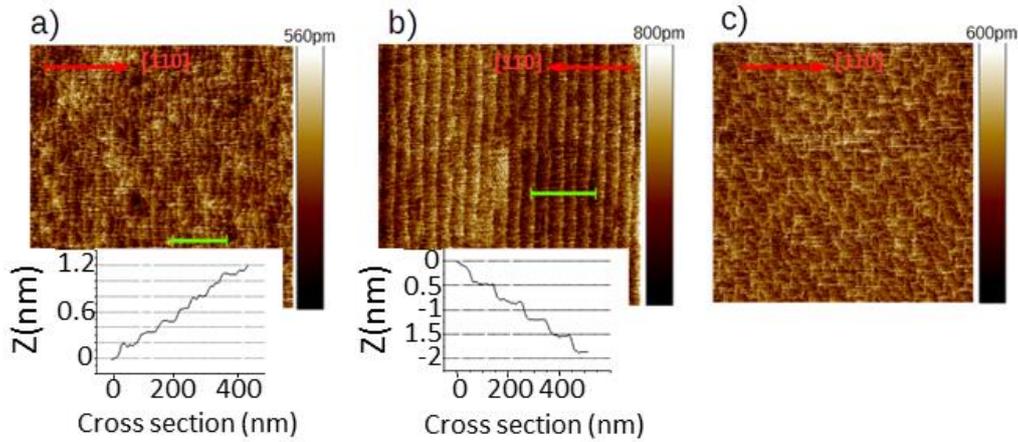

**Figure 2**: 2×2µm$^2$ AFM image of **(a)** Si(001) surface, 0.15°misoriented in [110], after $H_2$ annealing 900°C/10Torr. The surface is single-stepped with terraces width of ~50nm and step height of ~1.2 Å **(b)** Si(001) surface, 0.15°misoriented in [110], after $H_2$ annealing 900°C/600Torr. The surface is double-stepped with terraces width of ~100nm and step height of ~2.7 Å. **(c)** Si(001) surface, 0.12° misoriented in [100], after $H_2$ annealing 900°C/600Torr. The surface is single-stepped with dendritic steps oriented in <110> directions.

Then a 10 minutes $H_2$ annealing is performed at high pressure (near the atmospheric pressure) and the Si(001) surface changes to a double steps structure (figure 2b). The width of the terraces is measured around 100 nm which is consistent with the theoretical value for a double step terraces (L=$z_{DL}$/tan($\theta$)=103nm; where L is the terrace width; $\theta$=0.15° is the miscut angle; $z_{DL}$=2.72Å is the height of a DL step). The DL steps structure is also confirmed by the step height extraction from the AFM line profile (~2.7Å inset in figure 2b). For these conditions, the H chemical potentials is in the windows where the DFT calculations predict a high vacancies generation as well as a selectivity of the DVR over the DVL. Therefore, the DVR cross entirely the B-type terraces and this generates a nearly complete etching of the $S_B$ steps. This mechanism is expected to lead to the energetically unfavorable $D_A$ steps as seen by Brückner *et al* with 2° misoriented Si(001) wafer[36]. Nevertheless the AFM doesn't allow the atomic resolution necessary to confirm the step type ($D_A$ or $D_B$). One can also notice that a 10 min annealing is our optimal time to obtain DL steps. Shorter process times generate a surface with SL very similar to the one shown on Figure 2a. In other words, the $S_B$ steps etching process by the DVR is not complete with annealing times lower than 10 minutes.

A high pressure $H_2$ annealing of a Si(001) substrate with a 0.12° miscut in [100] direction was also tested. When the miscut direction slightly differs from the <110> azimuthal directions, each terrace boundaries are made of two types of step edge (with both $S_A$ and $S_B$ steps). After the $H_2$ annealing a sawtooth-like terraces are formed (figure 2c). This effect has already been observed by other authors[37-39]. This behavior is consistent with our mechanism. Actually, the dendritic single steps are made of DVR crossing the terraces in either [110] or [1-10] direction. In our experiments, it was not possible to achieve DL steps formation on wafers having a miscut direction different from <110>.

The last step is to evaluate the suitability of the silicon surface engineering strategy for III-V epitaxy co-integration. To do so, we have grown a GaAs layer on our optimized silicon surface (i.e. 0.15° in the [110] direction, surface corresponding to the figure 2b). We used trimethylgallium (TMGa) as group-III organometallic precursor and tertiarybutylarsine (TBAs) as group-V precursor. The carrier gas is hydrogen. The GaAs layer was grown with a standard two-step process: a 40 nm nucleation layer was first deposited at low temperature (LT 400-500°C), followed by 110 nm grown at higher temperature (HT 600-700°C)[40]. A V/III ratio below 10 is kept constant during all the process. The temperatures are monitored by optical pyrometers. The layer characterizations were done by AFM and scanning transmission electron microscopy (STEM) across thin lamellas prepared by focused ion beam (FIB).

Figure 3a –shows 2×2µm² AFM image of the GaAs nucleation layer surface after a 10s annealing at 600°C. The annealing is simply done to reveal the APBs since the Ga-Ga and As-As bonds can easily break up and evaporate leading to the typical V-groove shape topography of the APBs on the AFM pictures. The nucleation layer surface already presents a low APBs density due to prevalent double layer steps formation at the Si(100) surface. Few APBs, corresponding to the dark spots on the image, are still present. These remaining APBs originate from the few residual monoatomic islands that are still present after annealing of the Si(100) surface. Nevertheless after 110 nm growth of HT GaAs layer, the surface is completely APBs-free (Figure 3b). It results in a very low 0.8nm RMS roughness. Actually, the small APBs embodied in the nucleation layer have self-annihilated pairwise during the high temperature layer growth, as seen on the (110)-STEM cross-section of the layer (Figure 3c). In this picture most of the APBs seems to be perpendicular to the [110] axis. It is mainly due to the fact that we observe a projection of these planar defects in the cross section plane. Similar behavior was observed on GaP/Si[41] and GaAs/Si[42]. Beyond 70nm-thick the only defects observed, in all STEM cross-sections we did, are due to threading dislocations or stacking faults. No more APBs propagate in the layer as evidenced by the absence of V-groove shape defects on the surface observed by STEM and AFM. The impact of APBs on physical properties of GaAs /Si(001) has been reported in a previous paper[38].

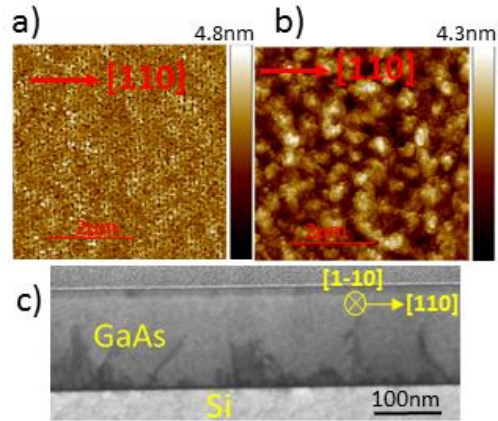

**Figure 3** : 5×5μm² AFM image of **(a)** the 40nm-thick GaAs nucleation layer, grown on optimized Si(001) wafer, after a 10s annealing at 600°C to reveal the APBs. The surface already has a very low APB density due to the prevalent double-steps on the Si substrate. **(b)** 110nm of GaAs grown at high temperature on the previous nucleation layer. The surface is APBs-free with a low RMS Roughness of 0,6nm. **(c)** (110)-STEM cross section of the GaAs layer. No APBs propagate until the surface.

In conclusion, the existence of a hydrogen chemical potential range where both the dimers vacancy density and the DVR selectivity over DVL are high enough to etch the $S_B$ steps is revealed by mean of DFT calculations. The latter are the key ingredients for double-layer steps formation on nominal Si(001) surface and allow to rationalize previous experimental reports[33,39]. These conditions are experimentally induced by using $H_2$ annealing at high temperatures (800-900°C) and high pressures (near atmospheric pressure) in a 300mm MOCVD epi-chamber. It results in the formation of a bilayer stepped surface, starting from a nominal Si(001) wafer slightly misoriented at 0.15° in the [110] direction. The efficiency of the silicon surface preparation is further validated by growing a thin (150nm) APBs-free GaAs layer on the optimized Si(001) substrate. The ability to get rid of APBs is a key point for a future monolithic integration of III-V nanostructures on a silicon platform.

This work has been partially supported by the LabEx Minos ANR-10-LABX-55-01 and the French "Recherches Technologiques de Base" (Basis Technological Research) and RENATECH programs. The DFT calculations were done using HPC resources from GENCI-CCRT (grant 2016-6107). The authors want to thank the CEA-Leti clean room staff, Stephane Puget for technical assistance on the MOCVD tool.